\documentclass[prb,twocolumn,amssymb,amsmath]{revtex4-1}

\usepackage{amsfonts}
\usepackage{bm}
\usepackage[usenames]{color}
\usepackage{graphicx}
\usepackage{epstopdf}

\newcommand{\gs}{{g_*}}
\newcommand{\es}{{\eta_*}}
\newcommand{\ds}{{\delta_*}}
\newcommand{\gz}{{g_0}}
\newcommand{\ez}{{\eta_0}}
\newcommand{\ees}{{\epsilon_*}}

\begin{document}

\title{Nematic fluctuations balancing the zoo of phases in half-filled quantum Hall systems}

\author{Andrej Mesaros$^{1}$, Michael J. Lawler$^{1,2,3}$ and Eun-Ah Kim$^{1,3}$}
\affiliation{$^{1}$Laboratory of Atomic and Solid State Physics, Cornell University, Ithaca, NY 14853, USA\\
$^{2}$Dept. of Physics, Applied Physics and Astronomy, Binghamton University, Vestal, NY 13850\\
$^{3}$Kavli Institute of Theoretical Physics, University of California, Santa Barbara, CA 93106}

\begin{abstract}
Half-filled Landau levels form a zoo of strongly correlated phases. These include non-Fermi liquids (NFL), fractional quantum Hall (FQH) states, nematic phases, and FQH nematic phases.  This diversity poses the question: what keeps the balance between the seemingly unrelated phases? The answer is elusive because the Halperin-Lee-Read (HLR) description that offers a natural departure point is inherent strongly coupled. But the observed nematic phases suggest nematic fluctuations play an important role. To study this possibility, we apply a recently formulated controlled double expansion approach in large-$N$ composite fermion flavors and small $\epsilon$ non-analytic bosonic action to the case with both gauge and nematic boson fluctuations. In the vicinity of a nematic quantum critical line (NQCL), we find that depending on the amount of screening of the gauge- and nematic-mediated interactions controlled by $\epsilon$'s, the RG flow points to all four mentioned correlated phases. When pairing preempts the nematic phase, NFL behavior is possible at temperatures above the pairing transition. We conclude by discussing measurements at low tilt angles which could reveal the stabilization of the FQH phase by nematic fluctuations.
\end{abstract}

\date{\today}

\maketitle


\section{Introduction}
\begin{figure}
  \centering
\includegraphics[width=0.42\textwidth]{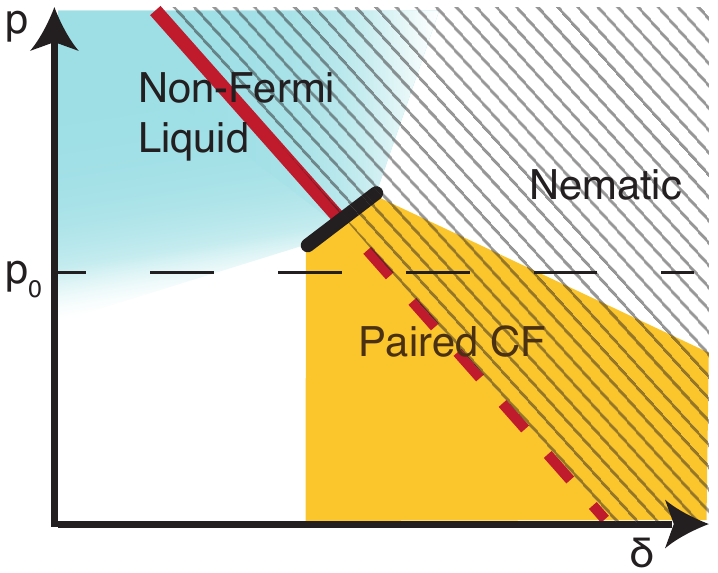}
\caption{Phase diagram in vicinity of quantum nematic transition occurring at critical isotropic pressure $p_c(\delta)$ [red line] as function of a measure $\delta$ of dominance of nematic-mediated over gauge-mediated bare interactions. The hatched region is nematically ordered. 
Along dashed red line the nematic transition is preempted by paired (composite-)fermion state (orange).
At ambient pressure ($p_0$) a non-Fermi liquid (blue) and gapless nematic (grey on right) are also present.
A continuous pairing transition occurs on black line.}
\label{fig:nqcp}
\end{figure}

Complexity of a phase diagram is a hallmark of strongly correlated systems and represents the rich physics of correlation.
It also challenges theoretical progress by making it hard for one to decide on the minimal model that will nevertheless faithfully represent the system of interest. 
Interestingly, upon a simple change of filling tuned by magnetic field, half-filled Landau levels switch through a zoo of exotic states that are commonly observed among strongly correlated materials.
Specifically, the $\nu=1/2$ state is one of the best established 
non-Fermi liquid states\cite{Halperin:1993p8143,Willett:1997p8742,Du:1993p8419,Du:1994p8746}, the $\nu=5/2$  state is the strongest candidate for a non-Abelian fractional quantum Hall state\cite{Blok:1990p6757,Nayak:1996hb,Fradkin:1998p8238,Willett:2013p8124,Willett:1987p8297,Tiemann:2012p8320} with $p+ip$ channel pairing of composite fermions\cite{Moore:1991p8317,Greiter:1992p8172} and the $\nu=9/2$ state is an electronic nematic state \cite{Fradkin:1999bl,Fradkin:2009vm,Lilly:1999p8109}. More recently, the $\nu=1/2$ state also attracted intense interest of the theory community \cite{Son:2015p8992,2015arXiv150505142M,Wang:2016ce}
as a possible gate into correlated topological surface states. 

Indeed the zoo of complex phases in the quantum Hall phase diagram have attracted many authors to view it as a paradigmatic place to explore quantum complexity.\cite{You:2014p8580,2016arXiv160201482Y,2010PhRvB..82h5102M,Cho:2014p8122,Abanin:2010p8579,2015PhRvB..92w5105K,2015PhRvB..91s5119J,2015PhRvB..92p5125B,Cho:2014jp}
Nevertheless, understanding the interplay of correlated states through a unified description remains an open question. In particular, the question of the mechanism of pairing in the fractional quantum Hall state remains open despite intense efforts and interest in the community.\cite{Nayak:2008dp} The clearest indication about pairing comes from numerical studies of interacting electrons in the $\nu=5/2$ ground state\cite{Rezayi:2000p8146,Lu:2010p8149,2015PhRvX...5b1004P}. Unfortunately theoretical understanding remains elusive, since fluctuations of the internal gauge field prevent the $p+ip$ pairing of composite fermions in the half-filled Landau level systems.\cite{Jain:1990p5667,Halperin:1993p8143,Bonesteel:1999p8234,Wang:2014p8425,Morinari:1998p8250}

Nematic fluctuations provide a clue to the question of pairing.
Phenomenologically 
not only the FQH $\nu=5/2$ state gives way to a nematic state with the gap closing under in-plane field\cite{Fradkin:2009vm,Pan:1999p8136,Lilly:1999p8137} it exhibits transport anisotropy before the gap closes\cite{You:2014p8580,Maciejko:2013p8119,Liu:2013p8581}. In particular a recent observation of transition between FQH $\nu=5/2$ state and a nematic induced by isotropic pressure\cite{2016NatPh..12..191S} strikingly demonstrates the proximity between the nematic state and the FQH $\nu=5/2$ state.
Interestingly, recently a number of theoretical works
have been establishing the idea that nematic fluctuations can enhance pairing
\cite{Metlitski:2010ws,Metlitski:2014p8103,Lederer:2015p8422,Schattner:2015p8990,2015arXiv151208523D,Li:2016kb}. Nevertheless little attention has been given to the role of putative nematic quantum critical fluctuations in forming the FQH $\nu=5/2$ state to date.
Here we study the role of nematic quantum critical fluctuations in the pairing of composite fermions assuming that a nematic quantum critical point can be accessed through a tuning parameter such as isotropic pressure (see Fig.~\ref{fig:nqcp}).
Moreover, as it is known that the filled Landau levels change the effective interactions\cite{Storni:2010p8184}, we envision a measure of dominance of nematic fluctuation to change with the changing of the Landau level (parametrized by $\delta$ in Fig.~\ref{fig:nqcp}). 
Hence we have a schematic phase space of Fig.~\ref{fig:nqcp} in mind, where the quantum critical value of pressure $p_c$ is changing with $\delta$ and defining a quantum critical line.

Specifically, we build on the recent progress in addressing the challenging problem of a Fermi surface coupled to massless fluctuations through a controlled perturbative renormalization group (RG) double expansion\cite{epsN_RG_Mross:2010p8159,Metlitski:2014p8103} and investigate the 
instabilities in systems in which both nematic and gauge fluctuations are present.

The rest of the paper is organized as follows: In Section~\ref{sec:model} we introduce the model and details of the RG procedure. Section~\ref{sec:phases} considers the resulting states, pairing and non-Fermi liquid behaviors at the NQCL. Behavior slightly away from the NQCL is considered in Section~\ref{sec:phasesR}. We close with discussion of applicability of our work, and summary of results and experimental predictions.

\section{Model}
\label{sec:model}
\begin{figure}
  \centering
\includegraphics[width=0.49\textwidth]{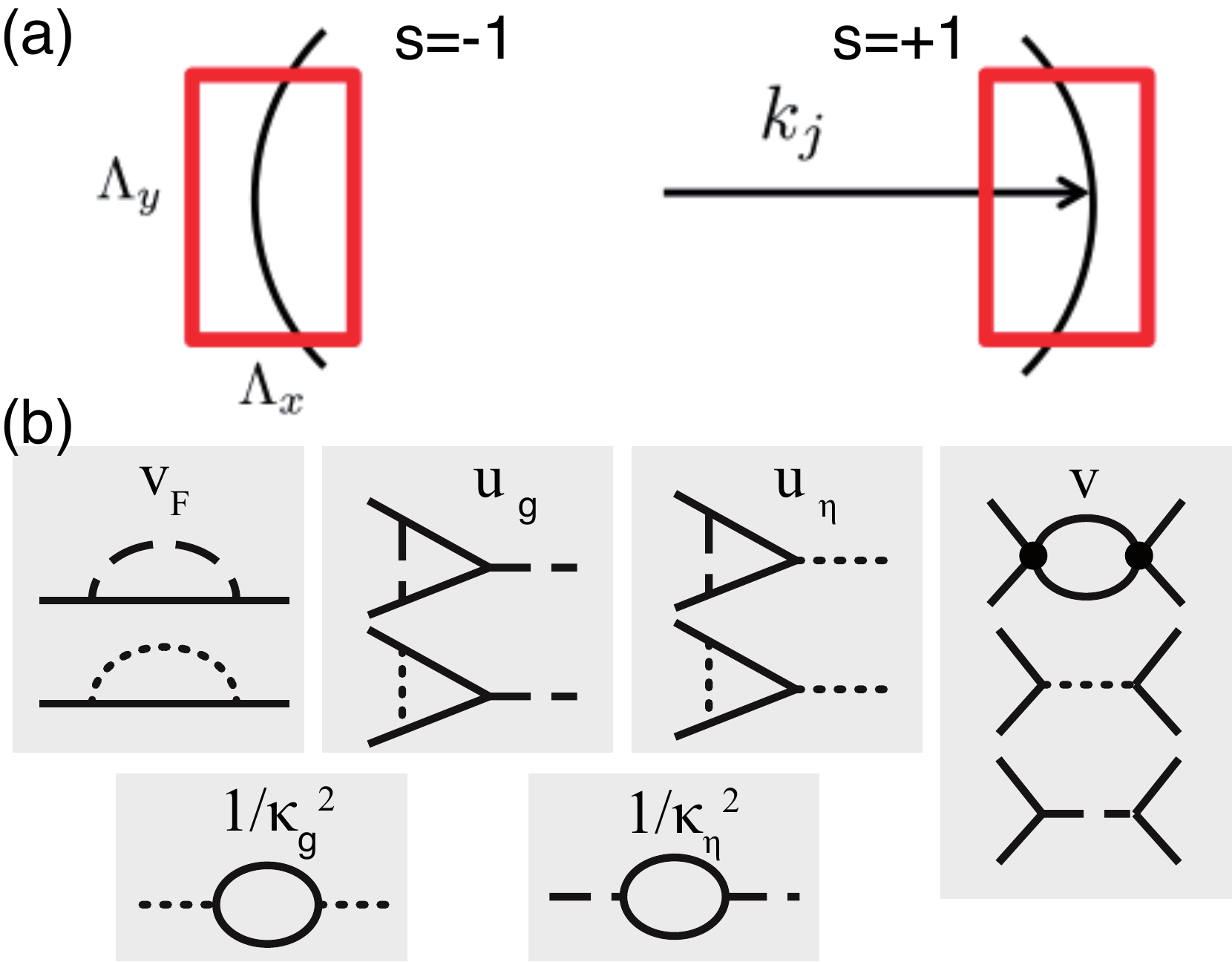}
\caption{Model and RG scheme. (a) Patch pair $j$, with corresponding Fermi momenta $\pm k_j$. (b) One-loop diagrams contributing to flow of coupling constants, with fermion propagator (full line), nematic propagator (dashed), gauge field propagator (dotted), and four-fermion interaction (dot).}
\label{fig:2}
\end{figure}
 
In order to study the interplay between the nematic quantum critical fluctuations and gauge fluctuations\cite{Halperin:1993p8143} in half-filled Landau levels, we extend the model in Ref.~\onlinecite{epsN_RG_Mross:2010p8159}. As in Ref.~\onlinecite{epsN_RG_Mross:2010p8159} we consider $N$ species of fermions and break up the Fermi surface of each species into independent patches\cite{2patchRG_Polchinski:1994}, i.e. decompose the $a$-th composite fermion field  $\psi^a(\tau,r)$ in two spatial dimensions $r$ and imaginary time $\tau$ with $a=1,\cdots, N$ into patch fields, i.e., 
\begin{equation}
\psi^a(\tau,r)=\sum_{j,s}\psi_j^{s,a}(\tau,r)e^{i sk_j r}
\end{equation}  
where the $j$-th patch pair is located at opposite Fermi momenta $s k_j$, $s=\pm 1$ (see Fig.~\ref{fig:2}a) assuming an inversion-symmetric Fermi surface. 
The action for the patch fermions are then given by 
\begin{equation}
    \label{eq:1f}
      S_j^f=\sum_{k,s,a}\bar{\psi}_j^{s,a}(k) D_s(k) \psi_j^{s,a}(k),
\end{equation}
where $D_s(k)\equiv -i\omega_k +v_F(s k_x+\frac{1}{2K}k^2_y)$ and $v_F$ is the Fermi velocity, $K$ is the local Fermi surface curvature, and $k\equiv(\omega_f,k_x,k_y)$ represent fermionic Matsubara frequency $\omega_f$ and two-dimensional momentum, the normalized sum is $\sum_{k}\equiv\frac{1}{\beta}\sum_{\omega_{f}}\int \frac{d^2k}{(2\pi)^2}$.

Ref.~\onlinecite{epsN_RG_Mross:2010p8159} considered fermions coupled to a single boson, controlling the RG expansion using
two small parameters: $1/N$, and the deviation\cite{epsRG_Nayak:1994p8158} of the boson's dynamic critical exponent from 2. 
For composite fermions coupled to nematic quantum critical fluctuations our fermions on the patch pair $j$ will be each coupled to two bosonic fields: the massless transverse component (in the direction of Fermi momentum, see Appendix~\ref{sec:hlr}) of the gauge field,\cite{2patchRG_Polchinski:1994} $\phi_{g}(\tau,r)$, and the nematic fluctuation, $\phi_{\eta}(\tau,r)$ which is massless at the NQCL. We then follow Ref. \onlinecite{Metlitski:2014p8103} and break the nematic and gauge fields into separate patch fields\cite{2patchRG_Polchinski:1994,Metlitski:2010ws} $\phi_{\eta}(\tau,r)$ and  $\phi_{g}(\tau,r)$. The bosonic action is then
\begin{align}
  \label{eq:1b}
    S_j^b&=\frac{1}{2\kappa_\eta^2}\sum_q |q_y|^{1+\epsilon_\eta}|\phi_{j,\eta}(q)|^2+\frac{r_0}{2}\sum_q\phi_{j,\eta}(q)|^2+\\\notag
  &+\frac{1}{2\kappa_g^2}\sum_q |q_y|^{1+\epsilon_g}|\phi_{j,g}(q)|^2 + \ldots
\end{align}
with $q\equiv(\omega_b,q_x,q_y)$ the bosonic Matsubara frequency and momentum variables, the
bare mass $r_0$ measures the distance to the NQCL from either side of the transition and "$\ldots$" represent all other irrelevant analytic terms that we will ignore. Here $\kappa_\eta$ and $\kappa_g$, the boson couplings for nematic and gauge boson, respectively, get renormalized under RG (see Fig 2b).
We retain control of the calculation for small enough nematic mass (see Appendix~\ref{sec:appR}), in a regime of strong fermion-nematic coupling which is complementary to the one accessed by Ref.\onlinecite{Lederer:2015p8422}.
The deviation of each boson's dynamic critical exponents from 2 represented by $\epsilon_\eta<1$ for the nematic fluctuation and $\epsilon_g<1$ for the gauge boson will control two double expansions together with $1/N$. Due to the non-analyticities in the action, these will not renormalize under RG\cite{epsN_RG_Mross:2010p8159}. 
Further, we will treat $\epsilon_\eta$ and $\epsilon_g$ as phenomenological parameters rather than view any particular value as ``physical". When coupled to fermions, these bosons mediate interaction between fermions. As filled Landau levels change the effective interaction between composite fermions,\cite{Scarola:2000p8228} in effect we anticipate the bare values of $\epsilon_\eta$ and $\epsilon_{g}$ to vary with the number of filled Landau levels and other external controls such as pressure. 

All together, the full effective action  $S_j$ for each patch pair $j$ is 
\begin{equation}
  \label{eq:1}
    S_j=  S_j^f+  S_j^b+  S_j^{int},
  \end{equation}
where $S_j^{int}$ represents the coupling between bosons and fermions
\begin{align}
  \label{eq:1int}
S_j^{int}&=\frac{u_\eta}{\sqrt{N}}\sum_{k,q}\phi_{j,\eta}(q)\sum_{s,a}\bar{\psi}_j^{s,a}(k+q)\psi_j^{s,a}(k)+\\\notag
  &+\frac{u_g}{\sqrt{N}}\sum_{k,q}\phi_{j,g}(q)\sum_{s,a}s\bar{\psi}_j^{s,a}(k+q)\psi_j^{s,a}(k),
\end{align}
with coupling constants $u_\eta$ and $u_g$ for the nematic-fermion and gauge-fermion interaction, respectively, being renormalized under RG (Fig.~\ref{fig:2}b). Note the difference in the sign of the coupling:\cite{Bonesteel:1999p8234}  nematic field couples to the density and hence the coupling is independent of the patch label, contrastingly the gauge field couples the current and hence the coupling has opposite signs on the two patches $s=\pm1$. 

Finally, our main goal is to investigate how the two critical couplings affect pairing of the composite fermions leading to the $\nu=5/2$ state. For this we
analyze fermion pairing instabilities by considering the residual composite fermion interaction in the BCS channel
\begin{equation}
  \label{eq:4}
S^{BCS}=-\frac{1}{4}\sum_{k,k',a}V^{\alpha\beta\gamma\delta}(k-k')\bar{\psi}_\alpha^{a}(k) \bar{\psi}_\beta^{a}(-k) \psi_\gamma^{a}(k') \psi_\delta^{a}(-k'),
\end{equation}
where we explicate spin indices, and use that in rotationally invariant system (true near enough to the NQCL) the interaction depends only on the angle of $q=k-k'$ with momenta $k$ and $k'$ taken on the Fermi surface. We consider the $S^{BCS}$ term without expanding in patch fermion species for efficiency. Inter-patch interactions it contains get renormalized when high-momentum bosons are integrated out and this form enables us to ignore the details of the patching procedure (Fig.~\ref{fig:2}b and Ref.\onlinecite{Metlitski:2014p8103}).

\section{RG flow and phase diagram on the NQCL}
\label{sec:phases}

Within the perturbative RG approach we consider the NQCL  Gaussian theory and the free-fermion fixed point, working in the zero-temperature limit at the NQCL. The
scaling which preserves the functional form of the fermionic propagator (Eq.~\eqref{eq:1f}) is 
\begin{align}
  \label{eq:2}
  &k_x\rightarrow t k_x\\\notag
  &k_y\rightarrow t^{1/2} k_y\\\notag
    &\omega\rightarrow t \omega,
  \end{align}
with $t=e^{-l}$, and $l$ being the RG scale. We set the same scalings for bosonic variables. To define the fermionic and bosonic modes to be integrated out, for every patch pair located at $\pm k_j$ we align the $x$-axis with $k_j$, fixing the patches as $|k_x|<\Lambda_x$, $|k_y|<\Lambda_y$, and then choose the high-energy fermion modes at $t\Lambda_x<|k_x|<\Lambda_x$ and bosonic ones at $\sqrt{t}\Lambda_y<|q_y|<\Lambda_y$. The fermionic modes at $\sqrt{t}\Lambda_y<|k_y|<\Lambda_y$ cross the Fermi surface and cannot be integrated out, so to preserve the patch aspect ratio with each RG step we relegate these modes to new patches.

The above RG method introduced in Ref. \onlinecite{Metlitski:2014p8103} is a hybrid between a two-patch scheme which focuses on interactions within a patch and a multipatch scheme which focuses on inter patch interactions. It merges the two schemes by being agnostic about how new patches are introduces at each RG step.  As such the method does not track information about the geometry of the Fermi surface.  
Although recent findings on importance of Fermi surface geometry was limited to Fermi surfaces in higher dimensions\cite{Mandal2015}, lack of a systematic scheme for introducing new patches may still harbor problems. Nevertheless, we proceed here assuming that there exists at least one well defined way to introduce new patches at each RG step.

The total action in Eq.~(\ref{eq:1}) has two dimensionless couplings at the NQCL: fermion-gauge coupling constant and fermion-nematic coupling constant, i.e.,
  \begin{align}
    \label{eq:3}
&g=\frac{u_g^2\kappa_g^2}{(2\pi)^2v_F\Lambda_y^{\epsilon_g}},\\\notag
&\eta=\frac{u_\eta^2\kappa_\eta^2}{(2\pi)^2v_F\Lambda_y^{\epsilon_\eta}},
\end{align}
respectively. Both couplings are relevant at our initial fixed point (Gaussian nematic and free fermion).
For the BCS instability, the coupling constants in Eq(\ref{eq:4})  $V^{\alpha\beta\gamma\delta}(k-k')$ in all spin-symmetric or spin-antisymmetric channels with fixed angular momentum are rendered indistinguishable within 1-loop RG and hence
may be labeled by single constant $V$. The corresponding dimensionless coupling constant is
\begin{equation}
  \label{eq:5}
  v=\frac{k_F}{2\pi v_F}V,
\end{equation}
where $v<0$$[v>0]$ is attraction[repulsion].
\begin{figure*}
  \centering
\includegraphics[width=0.95\textwidth]{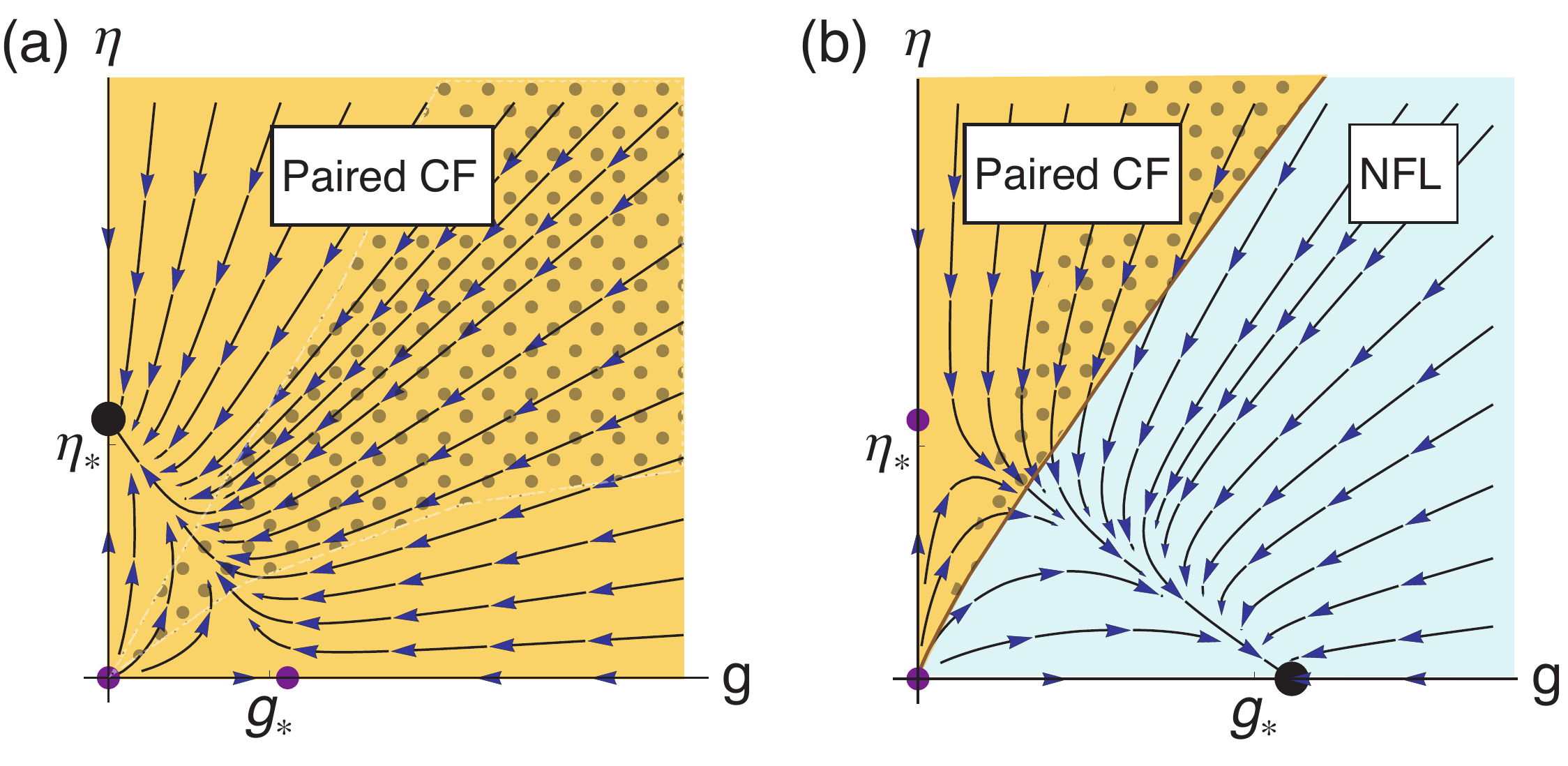}
\caption{RG flows at the NQCL projected onto the fermion-gauge ($g$) and fermion-nematic ($\eta$) coupling plane. Fixed points, stable (large dot) or unstable (small dot) in the plane are marked. In orange region the (composite-)fermions pair with vanishing bare attraction, i.e., a gapped FQH is formed. (a) The value $\gs/\es=0.7$ is representative for all $\epsilon_g<\epsilon_\eta$. In dotted orange area the NFL energy scale is higher than the pairing gap scale. (b) The choice $\gs/\es=1.44$ is representative for all $\epsilon_g>\epsilon_\eta$. The $(\gs,0)$ point is isotropic NFL (blue). At the boundary between the regions the pairing energy scale becomes zero. This phase diagram is robust away from nematic quantum critical line (see Fig.~\ref{fig:appptr}).}
\label{fig:1}
\end{figure*}

One-loop quantum corrections in our RG (see Fig.~\ref{fig:2}b) give the following flow equations for the Cooper pairing:
\begin{align}\notag
  &\frac{\textrm{d}v(l)}{\textrm{d}l}=-v(l)^2-f(l),\\
  &f(l)\equiv\eta(l)-g(l),\label{eq:11}
\end{align}
where we introduced the running coupling $f(l)$  to keep track of the competition between nematic and gauge fluctuations.
Positive $f$ promotes attraction in the BCS channel, and negative $f$ suppresses it.
Interestingly, within our theory not only can $f(l)$ have either sign but its sign can change during the RG flow.
The remaining couplings flow as:
\begin{align}
    \label{eq:7}
&\dot{g}=g\left(\frac{\epsilon_g}{2}-\frac{\eta}{N}-\frac{g}{N}\right)\\\notag
&\dot{\eta}=\eta\left(\frac{\epsilon_\eta}{2}-\frac{\eta}{N}-\frac{g}{N}\right)\\\notag
&\dot{v_F}=-v_F\left(\frac{\eta}{N}+\frac{g}{N}\right)\notag
\end{align}
where explicit $l$ dependence is dropped. The last equation shows that coupling to both the bosons enhances tendency towards non-Fermi liquid state, which is characterized by vanishing Fermi velocity $v_F$.

The two RG equations for the fermion-boson couplings $g,\eta$ in Eq.~(\ref{eq:7}) do not involve other couplings, so we start from the $g,\eta$ plane in which there are obvious fixed points: Beside the unstable free point $(g,\eta)=(0,0)$, there are $(g,\eta)=(\gs,0)$ and $(g,\eta)=(0,\es)$, where the defined numbers
\begin{align}
  \label{eq:45}
  &\gs\equiv\frac{N\epsilon_g}{2}\\\notag
  &\es\equiv\frac{N\epsilon_\eta}{2},
\end{align}
can take finite values in the double expansion $\epsilon_g,\epsilon_\eta\rightarrow 0$, $N\rightarrow\infty$. As the existence of fixed points is established, for simplicity in the following we set $N=1$ and consider the $\gs,\es\ll1$ limit.
In our approach, different experimental circumstances correspond to different bare values of running couplings $\eta_0$ and $g_0$, as well as to the balance between $\es$ and $\gs$ (i.e., $\epsilon_\eta$ and $\epsilon_g$, respectively). Since we take the physical value of bare pairing to be $v_0=0$, the pairing instabilities as well as non-Fermi liquid behavior are then fully determined by the values of $\eta_0, g_0, \es, \gs$.
The fine-tuned case $\epsilon_g=\epsilon_\eta$ is exceptional, and exhibits a line of fixed points connecting the two fixed points $(\gs,0)$ and $(0,\es)$ at the one-loop level (see Appendix~\ref{sec:appde}).
As the RG flows of $(g,\eta)$ and especially of the pairing coupling constant $v$ qualitatively differ depending on which one of the $\epsilon_g,\epsilon_\eta$ is larger, we analyze the two cases separately. 
For each case we infer the possible phases assuming the bare values of the fermion-boson couplings $(\gz,\ez)$ represent different experimental circumstances.  

In the case $\epsilon_g<\epsilon_\eta$ with the dynamic critical exponent of nematic boson being larger the only stable fixed point in the $(g,\eta)$ plane is $(g,\eta)=(0,\es)$ (see Fig.~\ref{fig:1}a). Obviously the fermions always pair (except when $\ez=0$ exactly), because $f(l)$ (see Eq.~(\ref{eq:11})) here flows from value $\ez-\gz$ to $\es>0$. Therefore eventually  $f(l)$ turns positive and stays so, giving a pairing instability even with $v_0=0$ to realize a gapped FQH state.
A remarkable consequence of this result is that the paired state is realized even in the limit in which the bare coupling of the fermions to the gauge fluctuations dominates over the bare coupling to the nematic fluctuations. Gauge fluctuations are no longer impeding pairing enough to push it to require a finite attractive interaction. Instead they only suppress the value of the pairing (FQH) gap estimated as $\Delta_P\sim\exp{(-l_P)}$ when a pairing instability $v=-\infty$ develops at a finite RG scale $l_P$. In the most extreme case of $\gz\rightarrow\infty, \ez\rightarrow0$ we obtained an analytic expression for the suppressed pairing gap in the limit 
 $\es-\gs\ll\gs$ to be (see Appendix~\ref{sec:appeta}):
\begin{equation}
  \label{eq:21}
  \Delta_P\sim \left(\frac{\ez}{\gz}\right)^{1/(\es-\gs)}\exp(-\pi/\sqrt{\es}).
\end{equation}
Although the NFL dictated by vanishing Fermi velocity in Eq.~(\ref{eq:7})) is unstable to infinitesimal pairing in the entire phase space of $(\gz,\ez)$ in this case, the NFL effects may be visible at temperatures above pairing $T_c$. This requires the energy scale associated with the NFL to be larger than the pairing gap scale, which occurs in the dotted region of Fig.~\ref{fig:1}a dictated by sufficiently large $(\gz+\ez)$ (see Appendix~\ref{sec:appnfl}).

\begin{figure}
  \centering
\includegraphics[width=0.45\textwidth]{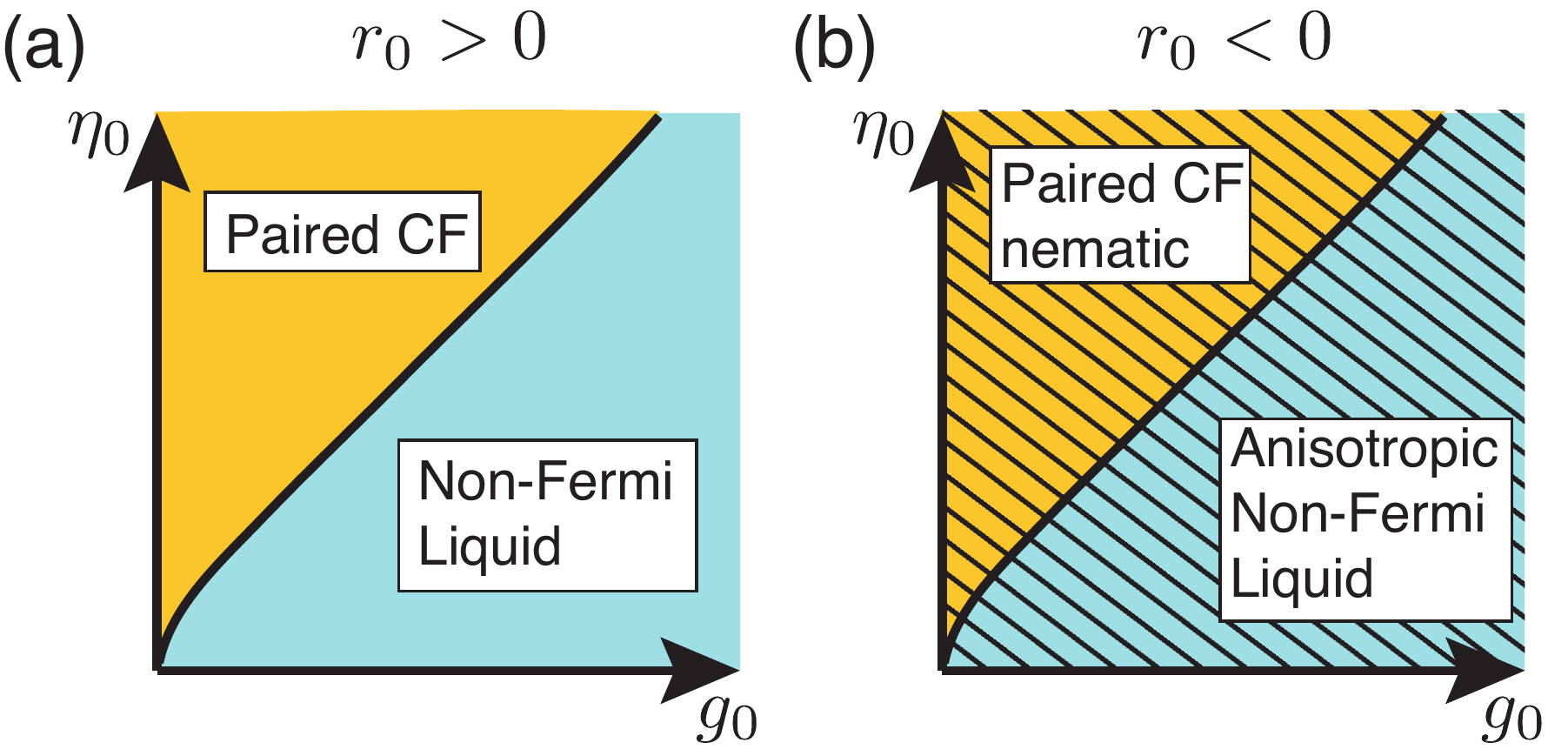}
\caption{Phases realized for different bare fermion-gauge ($\gz$) and fermion-nematic ($\ez$) couplings when the system is in vicinity of the nematic quantum transition. (a) In the disordered phase with positive nematic mass $r_0>0$ either a NFL (blue) or paired state (orange). (b) In the nematically ordered phase with negative nematic mass $r_0<0$ either anisotropic NFL (shaded blue) or gapped paired nematic state (shaded orange).}
\label{fig:4}
\end{figure}

In the case $\epsilon_g>\epsilon_\eta$ with the dynamic critical exponent of gauge boson being larger, there is a richer set of possibilities (see Fig.~\ref{fig:1}b). Namely, depending on the two bare boson-fermion coupling strengths we find either a stable non-Fermi liquid (blue region in Fig.~\ref{fig:1}b) or a paired state (orange in Fig.~\ref{fig:1}b). Moreover, we find the two phases in the $(\gz,\ez)$ plane are separated by a continuous
phase transition at the phase boundary given by 
\begin{equation}
  \label{eq:13}
(\ez-\gz)\ln\left(\frac{\ez}{\gz}\right)=(\gs-\es)\sqrt{\gs},
\end{equation}
for $\gs-\es\ll\gs$. Although the phase boundary in Fig.~\ref{fig:1}b needs to be obtained numerically in general (see Appendix~\ref{sec:appg}), a simple intuition can be gleaned from the beta function in Eq.~(\ref{eq:11}). When $\gz>\ez$ the function $f(l)$ stays negative throughout the RG flow and pairing requires above threshold strength of attractive bare interaction.
Therefore the fixed point $(\gs,0)$ controls the blue region of Fig.~\ref{fig:1}b. As Eq.~(\ref{eq:7}) dictates, the Fermi velocity $v_F$ flows to zero in this region resulting in a NFL phase driven by gauge fluctuations as in the original HLR model.\cite{Halperin:1993p8143,epsRG_Nayak:1994p8158}
On the other hand  pairing can occur as an infinitesimal instability for $\gz\ll \ez$ despite $(\gs,0)$ being the only stable fixed point. This is because $f(l)$ starts off positive for these bare values of couplings and the pairing instability can take over before $f(l)$ eventually turns negative.
The NFL effects may be visible again in the dotted region of the paired state (see Fig.~\ref{fig:1}b and after Eq.~\ref{eq:21}).
Furthermore, the continuity of the transition is evident by the fact that the pairing gap vanishes as $(\gz,\ez)$ approaches the phase boundary of Eq.~(\ref{eq:13}) with an analytic form we find for $\gs-\es\ll\gs$ 
\begin{equation}
\label{eq:14}
\Delta_P\sim x^{\frac{1}{2\sqrt{\gs}}}\left(\frac{\gz}{\ez}\right)^{1/\ds},
\end{equation}
where $x$ parametrizes a small distance in $(\gz,\ez)$ from the phase boundary (see Appendix~\ref{sec:appg}).

Overall we established that a composite Fermion system tuned to NQCL can be in one of two ground states: a paired state promoted by nematic fluctuations (orange regions in Fig.~\ref{fig:1}) or a stable NFL state governed by the gauge fluctuations (blue region in Fig.~\ref{fig:1}b).
If we associate the paired CF state with the $\nu=5/2$ FQH state our model indicates pairing in the $\nu=5/2$ FQH state is driven by the nematic fluctuations. Further associating the NFL state with the  $\nu=1/2$ NFL state, we are invited to postulate influence of nematic fluctuation to be weaker at lower Landau levels. If the degree of dominance between the two gapless bosons is varied with experimental conditions and the filling factor, and further the distance to the nematic phase can be varied with an external control such as isotropic pressure, we can now divide the NQCL into two parts as in Fig.~\ref{fig:nqcp}.

\section{Phases in the vicinity of NQCL}
\label{sec:phasesR}

Because accessing the quantum NQCL would require fine tuning, we now consider the effect of finite distance from the NQCL
with finite $r_0$ in Eq.~(\ref{eq:1b}). The positve mass $r_0>0$ will leave the system in the isotropic phase. But negative mass $r_0<0$ will drive the system into a nematic phase where nematic order parameter gains finite expectation value. However, the analysis of the fluctuation around this expectation value will closely follow the analysis of the nematic fluctuation in the isotropic phase. From here on we refer to the dimensionless coupling associated with the quadratic term in the action for the nematic fluctuation by $R \equiv r_0 \kappa_\eta^2/\Lambda_y^{1+\epsilon_\eta}$ which is always relevant.
Moreover a runaway flow of the nematic-fermion coupling $\eta$ takes the system to a strong-coupling regime outside the applicability of our methods when
 $\epsilon_g<\epsilon_\eta$.  
Nevertheless we can study the regime near NQCL by cutting the RG flows when $R$ reaches some limiting value. 

The nematic mass generally weakens the influence of nematic fluctuations and the RG equations now become
\begin{equation}
\label{eq:11R} 
f(l)\equiv\frac{\eta(l)}{1+R(l)}-g(l),
\end{equation}
and
\begin{align}
    \label{eq:7R}
&\dot{g}=g\left(\frac{\epsilon_g}{2}-\frac{\eta}{N}\frac{1}{1+R}-\frac{g}{N}\right)\\\notag
&\dot{\eta}=\eta\left(\frac{\epsilon_\eta}{2}-\frac{\eta}{N}\frac{1}{1+R}-\frac{g}{N}\right)\\\notag
&\dot{v_F}=-v_F\left(\frac{\eta}{N}\frac{1}{1+R}+\frac{g}{N}\right).\notag
\end{align}
Again we can establish a phase boundary between a paired state and NFL state 
in the $(\ez,\gz)$ phase diagram (see Appendix~\ref{sec:appR}). 
In the region of bare couplings where $\ez$ is sufficiently larger than $\gz$, the $f(l)$ starts out positive. If pairing instability takes over before $R(l)$ grows substantially the system will end up in a paired state. On the other hand, when $\ez\ll\gz$ $f(l)$ starts off negative and ultimately the rapid growth of $R$ ensures $f(l)\rightarrow-\gs$ as $l\rightarrow\infty$ leaving the system controlled by the gauge fluctuation without pairing. 
Now depending on the sign of $r_0$, the paired state and the NFL state each may be isotropic or nematic. Hence one can anticipate phase diagrams in Fig.~\ref{fig:4} with four distinct phases: isotropic paired CF, isotropic NFL (Fig.~\ref{fig:4}a), nematic paired CF, nematic NFL (Fig.~\ref{fig:4}b).
Indeed a systematic study of RG flows conforms to this anticipation (see Appendix~\ref{sec:appR}).
Hence within the regime of validity of our approaches, we see that the observation of nematic fluctuation driven pairing phase and a stable NFL phase we obtained at NQCL survives moving away from the NQCL line. The new facets introduced by considering the nematically ordered phase are possibilities of having anisotropic paired state and anisotropic NFL states (see Fig.~\ref{fig:nqcp}).

\section{Discussion and Conclusions}

To summarize, we used double expansion\cite{epsN_RG_Mross:2010p8159} in boson dynamic exponents\cite{epsRG_Nayak:1994p8158} and number of fermion species\cite{2patchRG_Polchinski:1994} to study NQCL and its vicinity in composite Fermi fluid. This approach has several issues including relying on non-analytic bosonic actions and an incompletely specified RG prescription. Nevertheless, we found the interplay between the gauge fluctuations and nematic fluctuations to account for the entire zoo of correlated states observed in half-filled Landau levels. To start with we capture the  NFL state at $\nu=1/2$,
FQH state at $\nu=5/2$ (paired CF state) and gapless nematic state at $\nu=9/2$. Moreover, the gapped FQH nematic observed under tilted filed experiment\cite{Liu:2013p8581} naturally appears on the nematic ordered side of the NQCL with the pairing driven by nematic fluctuation. Finally, recent observation of transition between a FQH state and a nematic state driven by isotropic pressure suggests the NQCL we envision in Fig.~\ref{fig:nqcp} can be accessed using pressure\cite{2016NatPh..12..191S}.

The key new insight that emerges from our 
result is that the pairing of CF's in $\nu=5/2$ systems can be driven by nematic fluctuations in the vicinity of NQCL.
Therefore we predict the magnitude of FQH gap in $\nu=5/2$ to be correlated with the nematic fluctuations which can be quantified through measuring nematic susceptibility. To achieve this, one possibility is to measure the nematic susceptibility in $\nu=5/2$ states by studying nematicity as a function of small tilt-angles. Then our results predict the nematic susceptibility so-measured to be monotonically correlated with the size of the FQH gap at zero tilt-angle. 

\vspace{5mm}
{\bf \noindent Acknowledgement} We thank Jim Eisenstein, Eduardo Fradkin, Michael Manfra, David Mross, Michael Mulligan and Senthil Todadri for helpful discussions. AM and E-AK are supported by the U.S. Department of Energy, Office of Basic Energy Sciences, Division of Materials Science and Engineering under Award DE-SC0010313. E-AK acknowledges support through Simons Fellowship for Theoretical Physics. E-AK and MJL are grateful to the hospitality of Kavli Institute of Theoretical Physics (KITP) during the completion of this work. E-AK and MJL acknowledge support by the National Science Foundation under Grant No. NSF PHY11-25915 through KITP.

\appendix
\section{From HLR gauge field to a scalar field}
\label{sec:hlr}
We briefly review the HLR model\cite{Halperin:1993p8143} and how it leads to the action for gauge field in Eq.~(\ref{eq:1b}) and its coupling to fermions in Eq.~(\ref{eq:1int}). The central insight of HLR is to attach two flux quanta of a $U(1)$ gauge field $\vec{a}$ to each electron which creates a composite fermion (CF) denoted by field $\psi(\vec{r})$, as expressed by the constraint
\begin{equation}
  \label{eq:6}
\vec\nabla\times\vec{a}(\vec{r})=2(2\pi)\psi ^\dagger (\vec{r})\psi(\vec{r}),
\end{equation}
where the CF density on the right-hand side equals the original electron density. An HLR action with $\tau$ the imaginary time therefore contains a Chern-Simons term for the gauge field $a_\mu=(i a_\tau,\vec{a})$ which provides the flux attachment as is obvious when the $a_\tau$ component is integrated out to recover Eq.~\eqref{eq:6}:
\begin{align}
  \label{eq:8}
  &S_{CF+gauge}\!=\!\int\!\!\textrm{d}\tau\textrm{d}^2\vec{x}\;\psi^\dagger \left[\partial_\tau-i a_\tau +E(-i\vec{\nabla}+e\vec{A}-\vec{a})\right]\psi,\\
    &S_{CS}\!=\!\int\!\!\textrm{d}\tau \textrm{d}^2\vec{x}\;\frac{1}{2(4\pi)}\epsilon_{\mu\nu\lambda}a_\mu\partial_\nu a_\lambda,
\end{align}
where $\vec{A}$ is electromagnetic potential and $E(\vec{k})$ the electron dispersion. In half-filled Landau levels, the attached $a_\mu$ gauge flux in mean-field approximation exactly cancels the external magnetic flux leaving the CF free, however both the fluctuations of $a_\mu$ and the interactions between the CF particles cannot be ignored and it is advantageous to treat them together. The interaction between CF particles is effective and therefore considered to have varying range
\begin{equation}
  \label{eq:36}
  S_{CF int}\!=\!\int\!\!\textrm{d}\tau \textrm{d}^2\vec{x}\textrm{d}^2\vec{y}\frac{U}{|\vec{x}-\vec{y}|^{1+\epsilon}}\psi^\dagger (\tau,\vec{x}) \psi(\tau,\vec{x}) \psi^\dagger (\tau,\vec{y}) \psi(\tau,\vec{y}),
\end{equation}
from Coulomb for $\epsilon=0$ to short-range as $\epsilon\rightarrow 1$, giving the full HFL action $S_{HLR}=S_{CF+gauge}+S_{CS}+S_{CF int}$. The CF density in quartic term $S_{CF int}$ allows one to rewrite it exactly as a purely gauge field quadratic term using the constraint Eq.~\eqref{eq:6} which also implies that only the transverse component of the gauge field $a_T(\tau,\vec{k})\equiv (\hat{z}\times\hat{k})\cdot\vec{a}(\tau,\vec{k})$ at momentum $\vec{k}$ appears:
\begin{equation}
  \label{eq:37}
  S'_{CF int}\!=\!\int\!\!\textrm{d}\tau \textrm{d}^2\vec{k}\;\frac{U}{|\vec{k}|^{1-\epsilon}}\frac{1}{(4\pi)^2}|\vec{k}|^2a_T(\tau,\vec{k})a_T(\tau,-\vec{k}),
\end{equation}
where we dropped an $\epsilon$-dependent normalization to obtain the term in our action Eq.~(\ref{eq:1b}) where $a_T$ is relabeled to $\phi_{j,g}$ after restriction of its momenta determined by patch-pair $j$ (below Eq.~(\ref{eq:2})). Through transformation from $S_{CF int}$ to  $S'_{CF int}$ it was recognized that a non-Fermi liquid fixed point can be accessed in a perturbative expansion of $\epsilon$.\cite{epsRG_Nayak:1994p8158}

The CFs couple strongly to the transverse gauge component $a_T$ due to the scaling transformation (Eq.~(\ref{eq:2})) which is chosen to preserve the Fermi surface at patch-pairs $j$ as they scale towards Fermi point $\pm\vec{k}_{F,j}=\pm k_F\hat{x}_j$.\cite{2patchRG_Polchinski:1994} Since for a circular Fermi surface the CF current in patch $j$ gets directed along $x$-axis, the expansion of CF-gauge coupling term $S_{CF+gauge}$ in Eq.~(\ref{eq:8}) has the lowest order term in powers of gauge field and derivatives $v_F a_{x_j}({\psi_j^+}^\dagger \psi^+_j-{\psi_j^-}^\dagger \psi^-_j)$ (note that the fermion species index is suppressed). On the other hand the patches scale such that their aspect ratio remains $\Lambda_{x_j}\sim\Lambda_{y_j}^2/k_F\ll\Lambda_{y_j}$ (see below Eq.~(\ref{eq:2})) so that in RG transformation of patch-pair $j$ the relevant high-energy gauge modes have momenta $q_{x_j}\ll q_{y_j}$. Therefore in every patch-pair $j$ the gauge component that couples dominantly to CF's is transverse, i.e. $a_{x_j}$ with momentum $q_{y_j}$.

\section{RG diagram for $\epsilon_g=\epsilon_\eta$}
\label{sec:appde}

In this situation the flow equations (\ref{eq:7}) lead to flow along rays through the origin:
\begin{equation}
  \label{eq:18}
  \frac{\eta}{g}=\frac{\ez}{\gz},
\end{equation}
and there is a line of fixed points $(g_*,\eta_*)$ satisfying:
\begin{equation}
  \label{eq:10}
  g_*+\eta_*=\ees,
\end{equation}
where we defined $\ees\equiv N\epsilon_g/2=N\epsilon_\eta/2$.

On the NQCL, the pairing function $f(l)$ changes from $\ez-\gz$ to $(\ez-\gz)\frac{\ees}{\ez+\gz}$, so there is an infinitesimal pairing instability only for $\ez>\gz$.
Right at the line $\ez=\gz$ there is BCS behavior which is not expected to be generic beyond one-loop.

The NFL energy scale dominates over the pairing scale inside the strip $\ez-\gz\ll(\ez-\gz)^2$ in the $\ez>\gz$ region, while the converse is found for $\ez-\gz\gg(\ez-\gz)^2$ (see Appendix~\ref{sec:appnfl}).

\section{Pairing for $\epsilon_g<\epsilon_\eta$}
\label{sec:appeta}

We can find analytic approximations for the flow of pairing in the limit of $\gs$ and $\es$ being similar:
\begin{equation}
  \label{eq:23}
  |\ds|\ll\gs,\text{ where } \ds\equiv\es-\gs.
\end{equation}
This limit is generally convenient as it provides a separation of scales in the RG flow of fermion-boson couplings: the flow of $(g,\eta)$ is near $(\gz,\ez)$ for $l\ll 1/\es$, near the line in Eq.~(\ref{eq:10}) for $1/\es\ll l\ll 1/|\ds|$, and near the fixed point for $l\gg 1/|\ds|$.
The separation of scales follows from the analytic solution of RG flow for fermion-boson couplings $g,\eta$ on the NQCL:
\begin{align}
  \label{eq:9}
&\eta(l)=\frac{\ez \exp(\es l)}{1+h(l)-h(0)},
&h(l)=\frac{\ez}{\es}e^{\es l}+\frac{\gz}{\gs}e^{\gs l}.
\end{align}

For the case $\gs<\es$ we identify several regimes for the pairing gap scale in the region $\ez<\es$. With the separation of scales, the $f(l)$ remains approximately constant for scales $l\ll1/\es$, and in the least favorable case for pairing, $\ez\ll\gz$, the value is $f(l)\approx-\gz$. It is useful to analyze in general the flow of pairing $v$ (Eq.~(\ref{eq:7})) when $f(l)$ is constant. The outcome depends drastically on the sign of the constant. For $f(l)=-D^2$, $D>0$, the flow is
\begin{equation}
  \label{eq:16}
  v_D(l)=D\frac{v_{init}+D+(v_{init}-D)\exp(-2Dl)}{v_{init}+D-(v_{init}-D)\exp(-2Dl)},
\end{equation}
with attractive fixed point at $v_{init}=D$ and repulsive one at $v_{init}=-D$. Only if $v_{init}<-D$ there is a pairing instability $v\rightarrow-\infty$ reached at $l_{-D}=\frac{1}{2D}\ln\left(\frac{D+|v_{init}|}{D-|v_{init}|}\right)$. Therefore the pairing interaction $v$ in the beginning part of the flow ($l\ll1/\es$) settles at the repulsive value $v_+=+\sqrt{\gz}$ (we assume $v_0=0$). At the scale
\begin{equation}
  \label{eq:20}
l_T=\frac{1}{|\ds|}\ln\left(\frac{\gz}{\ez}\right)
\end{equation}
$f(l)$ becomes positive, and most of the ensuing flow has $f(l)\approx\es$. So here we can use the solution to the flow with $f(l)$ being positive and constant, and the initial condition being $v_{init}=v_+=+\sqrt{\gz}$. The solution to the flow of $v$ given $f(l)=C^2$, $C>0$, is
\begin{equation}
  \label{eq:15}
  v_C(l)=C\tan[-C l+\arctan(v_{init}/C)],
\end{equation}
with pairing instability $v\rightarrow-\infty$ reached at $l_{C+}=\frac{\pi}{C}$ for $v_{init}\gg C$, at $l_{C-}=\frac{1}{|v_{init}|}$ for $v_{init}\ll -C$ (weak coupling BCS case), and at $l_{C0}=\frac{\pi}{2C}$ for $|v_{init}|\ll C$. (Note that drastically different from Eq.~(\ref{eq:16}), there is always a pairing instability.) Applying Eq.~(\ref{eq:15}) therefore gives for the unfavorable case $\gz\gg\es$ the pairing scale $l'_P=\pi/\sqrt{\es}$. The total pairing scale is then $l_T+l'_P$, leading to Eq.~(\ref{eq:21}).

To estimate the pairing scale when the gauge-fermion bare coupling diminishes, for example when $(\gz,\ez)$ is close to the unstable fixed point $(\gs,0)$, we solve a differential equation obtained in various approximations to the flow equation of $v$ (Eq.~(\ref{eq:7})). Let us consider the function $f$ of the form
\begin{equation}
  \label{eq:30}
  f(l)\equiv a\exp(k l)-b,
\end{equation}
with $a,b,k$ positive constants. With the substitutions $\bar{v(l)}=v(l)/k$, $\bar{f(l)}=(1/k^2)(f(l)+b)$, we obtain the flow equation
\begin{equation}
  \label{eq:25}
  \bar{f}\frac{d\bar{v}}{d\bar{f}}=-(\bar{v}^2+\bar{f}-b/k^2).
\end{equation}
The solution takes the functional form $\bar{v}=h(\bar{f},b/k^2,C)$, with
\begin{widetext}
\begin{equation}
  \label{eq:26}
  h(x,y,c)\equiv \frac{\sqrt{x}}{2}\frac{C\Gamma(1 - 2\sqrt{y})(J_{-2\sqrt{y}-1}(2\sqrt{x}) - 
              J_{1 - 2\sqrt{y}}(2\sqrt{x})) + 
     \Gamma(2\sqrt{y} + 1)
           (J_{2\sqrt{y} - 1}(2\sqrt{x}) - 
        J_{2\sqrt{y} + 1}(2\sqrt{x}))}{
   C\Gamma(1 - 2\sqrt{y}) J_{-2\sqrt{y}}(2\sqrt{x}) + 
         \Gamma(2\sqrt{y} + 1) J_{2\sqrt{y}}(2\sqrt{x})}.
\end{equation}
\end{widetext}
The integration constant $C$ is fixed by initial conditions $v(0)=0$, $\bar{f}(0)=a$, giving $h(a/k^2,b/k^2,C)\equiv0$ which is easily solved to obtain the parameter-dependent value $C(a/k^2,b/k^2)$. Labeling the denominator of $h$ in Eq.~(\ref{eq:26}) by $\chi$, the pairing instability $v\rightarrow-\infty$ occurs at scale $l_P$ when $\chi\equiv0$, which gives the implicit equation
\begin{equation}
    \label{eq:27}
    \chi\left(\frac{a}{k^2}e^{k l_P},\frac{b}{k^2},C\left(\frac{a}{k^2},\frac{b}{k^2}\right)\right)=0.
\end{equation}
We write $l_P$ in the form
\begin{equation}
  \label{eq:28}
l_P\equiv\frac{1}{k}\ln\left(\frac{k^2}{a}z_P\right),  
\end{equation}
where $z_P$ in principle depends on $a,b,k$. This corresponds to saying that $\bar{f}(l_P)\simeq1$.

\begin{figure}
  \centering
\includegraphics[width=0.49\textwidth]{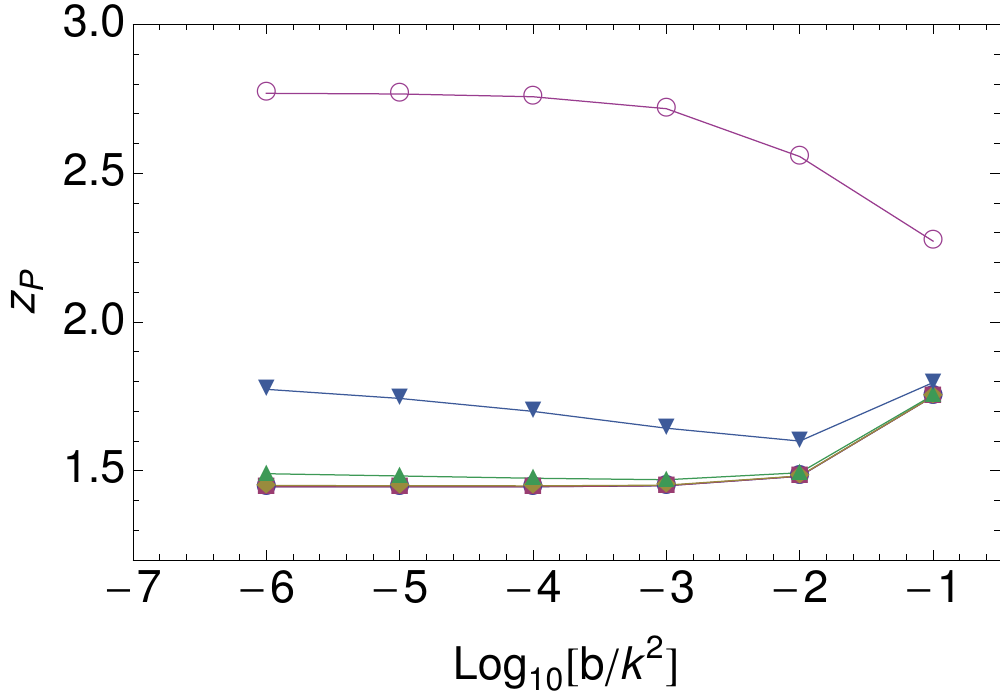}
\caption{Pairing scale factor $z_P$ (Eq.~(\ref{eq:28})) dependence on the parameters $a/k^2$ and $b/k^2$ (Eq.~(\ref{eq:30})). The curves from bottom to top are for $a/k^2=10^{-6}, 10^{-5}...10^{-1}$.}
\label{fig:app3}
\end{figure}

We numerically find that $z_P$ itself depends very weakly on the parameters and is practically constant of order 1 up to $a/k^2,b/k^2\lesssim0.1$ (see Fig.~\ref{fig:app3}).

Returning to the physical problem of intermediate and weak bare fermion-gauge coupling which is still much stronger than bare fermion-nematic coupling, on the NQCL the function $f(l)$ can in general be rewritten to emphasize the dependence on $\ds$:
\begin{align}
  \label{eq:29}
  &f(l)=\frac{\ez\frac{\gs}{\gz} \exp(\ds l)-\gs}{1+X \exp(-\gs l)+\frac{\gs\ez}{\gz\es} \exp(\ds l)},\\\notag
  &X\equiv \frac{\gs}{\gz}\left(1-\frac{\ez}{\es}-\frac{\gz}{\gs}\right)
\end{align}
where $X$ measures the distance from the line connecting the fixed points (see Eq.~(\ref{eq:10})). The flow of $v$ becomes analytically tractable when $f$ reduces to the form in Eq.~(\ref{eq:30}). We can therefore consider the example case of $(\gz,\ez)$ close to the unstable fixed point $(\gs,0)$ by its behavior on the line $X=0$. Here the denominator of $f$ is approximately 1 on scales $l\ll l_v\equiv(1/\ds)\ln(\es/\ez-1)$. Using $X=0$ to eliminate $\gs/\gz$ we have the tractable form $f(l)\approx\frac{\ez}{1-\ez/\es}\exp(\ds l)-\gs$. The consistency condition $l_P\ll l_v$ reduces to $\ez/\es\ll1$. Using this, the result $l_P=(1/\ds)\ln(\ds^2/\ez)$ follows from Eq.~(\ref{eq:28}) and the pairing gap is
\begin{equation}
  \label{eq:22}
\Delta_P\sim \left(\frac{\ez}{\ds^2}\right)^{1/\ds}.
\end{equation}
Next we consider an even weaker gauge-fermion bare coupling $\gz$, more precisely the regime $\frac{\ez}{\es}-\frac{\gz}{\gs}\ll1$, further assuming that $l_P\ll1/\ds$ which makes the numerator of $f(l)$ constant and the $X$ term in denominator dominant (see Eq.~(\ref{eq:29})). The result $l_P=(1/\gs)\ln(\gs^2/(\ez-\gz))$ follows, under the constraint $\ez>\gz$ and $l_P\ll1/\ds$. The latter condition can be rewritten as $\ez/\es+\gz/\gs\ll1$ and $\ez-\gz\gg\gs^2\exp(-\gs/\ds)$, while the gap becomes a stronger power-law:
\begin{equation}
  \label{eq:24}
\Delta_P\sim \left(\frac{\ez-\gz}{\es^2}\right)^{1/\es}.
\end{equation}

Finally, we note that setting $\gs\ll\es$, i.e., considering a phase diagram far from the case in Eq.~(\ref{eq:23}), one expects the results to reduce to the ones in Ref.\onlinecite{Metlitski:2014p8103} having fermion-nematic coupling only. In the $\ez\ll\es$ region of phase diagram this is indeed true. Taking $\ez\ll\es$, in three regimes $\gz\ll\gs\ll\es$, $\gz=\gs\ll\es$ and $\gs\ll\gz\ll\es$ we find that
\begin{equation}
  \label{eq:31}
f(l)\approx\ez\exp(\es l)-\gz,
\end{equation}
which leads to the result $l_P=(1/\es)\ln((\es^2/\ez))$, due to $z_P\approx1$ (see \ref{eq:28}).

\section{Pairing transition line for $\epsilon_g>\epsilon_\eta$}
\label{sec:appg}
\begin{figure}
  \centering
\includegraphics[width=0.49\textwidth]{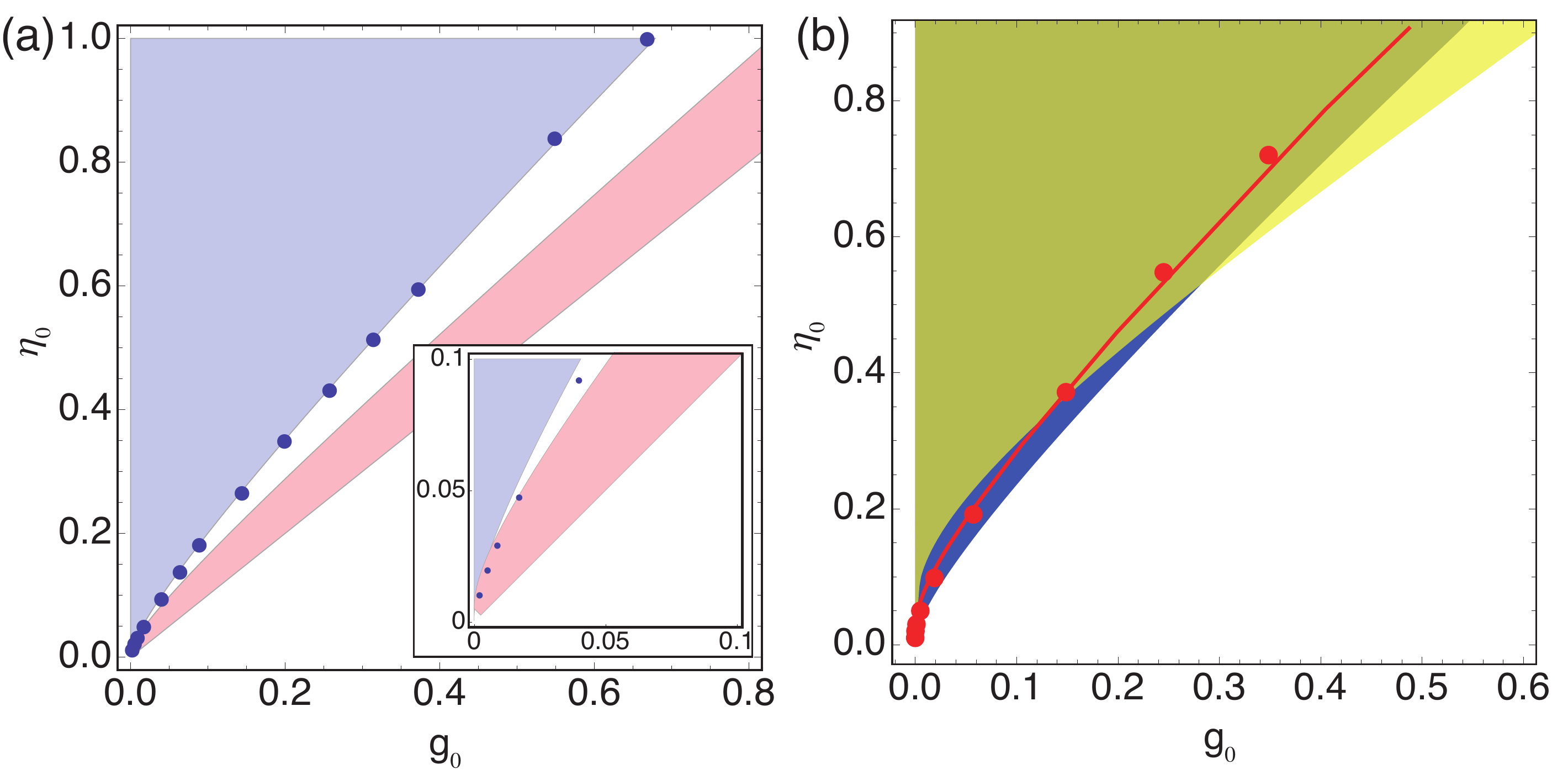}
\caption{Line of continuous transition to paired state in the plane of bare couplings $(\gz,\ez)$, for $\epsilon_g>\epsilon_\eta$. (a) On the NQCL, with $\gs=1.3,\es=0.9$. Blue dots are obtained from numerical solutions to RG flow by having the $v(l)$ coupling diverge near largest available scale $l$. Boundary of blue shading is approximate expression Eq.~(\ref{eq:43}), and upper boundary of red shading is Eq.~(\ref{eq:13}); see discussion in Appendix~\ref{sec:appg}. Inset: Zoom-in at small bare couplings. (b) Away from NQCL, $R_0=0.05$, and $\gs=1.3,\es=0.8$. Red dots are obtained from numerical solutions to RG flow by having the $v(l)$ coupling diverge near largest available scale $l$. Red line is obtained by equating $l'_T$ (scale at which pairing function $f(l)$ starts repressing pairing tendency), obtained numerically from Eq.~(\ref{eq:40}), to the approximation $l_P=\pi/(2\sqrt{\ez-\gz})$ (see discussion in Appendix~\ref{sec:appg}). Boundary of blue and yellow shadings use the two approximate expressions for $l'_T$ in Eq.~(\ref{eq:41}), respectively. Note that blue corresponds to the $l_T$ with $R_0=0$, while yellow matches the slope near the origin better.}
\label{fig:app2opt}
\end{figure}

To derive the expression for the pairing transition line in the plane of boson-fermion couplings, we focus on the region $\ez>\gz$. Assuming the separation of scales (see Eq.~(\ref{eq:23})), we can approximately replace the $f(l)$ by the positive constant $\ez-\gz$ in the first part of the flow, and the negative constant $-\gs$ in the second part of the flow. The vanishing pairing gap at the transition line implies that the pairing scale found in the second part of flow is $l'_P\rightarrow\infty$, which is a condition we use to connect the solutions in the two parts of the flow. The $f(l)$ becomes negative at
\begin{equation}
  \label{eq:12}
  l_T=\frac{1}{|\ds|}\ln\left(\frac{\ez}{\gz}\right),
\end{equation}
and once it does it quickly approaches the constant value $f=-\gs$. We can use this in the solution of Eq.~(\ref{eq:16}) for the second part of the flow, except that the initial condition $v_{init}$, being the value of $v$ when the flow entered the regime $f\approx-\gs$, is still unknown. Our demand that $l'_P\rightarrow\infty$ occurs if $v_{init}$ is just below $-\sqrt{\gs}$ (see Eq.~(\ref{eq:16})). So we can set $v_{init}\equiv-\sqrt{\gs}$, and use this as a condition for the first part of the flow. The $v_{init}$ can be estimated as the value $v(l_T)$, while the latter can be estimated by using the first part of flow, where $f\approx\ez-\gz\equiv C^2$. Using Eq.~(\ref{eq:15}) therefore gives the implicit equation that corresponds to vanishing pairing gap:
\begin{equation}
  \label{eq:17}
v_{C}(l_T)\equiv-\sqrt{\gs}.
\end{equation}
Assuming $C l_T\ll 1$ the tangent can be approximated and Eq.~(\ref{eq:17}) gives Eq.~(\ref{eq:13}) of main text, which is consistent with $C l_T\ll 1$ as long as $\gz,\ez$ are not orders of magnitude larger than the values of $\gs,\es$.

We tested the prediction in Eq.~(\ref{eq:13}) by numerically solving the flow, see Fig.~\ref{fig:app2opt}a. There is good agreement in the considered regime $|\ds|\ll\gs$; $\gz,\ez\ll\gs$, however we note that there is excellent agreement with the line
\begin{equation}
  \label{eq:43}
  \sqrt{\ez-\gz}\ln\left(\frac{\ez}{\gz}\right)=\pi|\ds|/2
\end{equation}
in a wider parameter range. This is a noteworthy property of the numerical experiment: it is challenging to numerically observe a divergence $v(l)\rightarrow-\infty$ at large values of $l\equiv l_P$.  Deep in the considered regime $|\ds|\ll\gs$; $\gz,\ez\ll\gs$ we could identify flows where $v$ upon entering into second part of flow (see before Eq.~\ref{eq:12}) hovers at a fixed negative value for long stretches of $l$ before diverging. This is precisely the expected behavior in the approximation of $f(l)$ being constant in two parts of the flow, in which case $v$ in the second part starts out just below its unstable fixed point $-\sqrt{\gs}$. However, with a given numerical precision and a wider range of initial conditions, it becomes hard to tune $\gz,\ez$ such that this second part of flow of $v$ is realized. Instead, one easily identifies the values $\gz,\ez$ for which $v(l)$ develops a divergence at a relatively small $l\equiv l_P$ while $f(l)$ is still not too negative and therefore $l$ is close to value $l_T$ where $f(l)$ changes sign. That kind of numerical identification of transition point corresponds to equating $l_P$ with $l_T$. Identifying $l_P=\pi/(2\sqrt{\ez-\gz})$, which is particularly good approximation for slower flows when $\ez,\gz$ are comparable to $\es,\gs$, the quoted expression Eq.~(\ref{eq:43}) follows directly. Of course, $l_T$ is finite so the condition $l_P=l_T$ in principle does not allow $l_P\rightarrow\infty$ and $\Delta_P\rightarrow0$. However, the similarity of curves in Fig.~\ref{fig:app2opt}a shows that the error in our numerical identification of $l_P$ is small.

The pairing energy scale in the main text is determined using the total pairing scale $l_T+l'_P$.

\section{Flow of Fermi velocity}
\label{sec:appnfl}
\begin{figure}
  \centering
\includegraphics[width=0.49\textwidth]{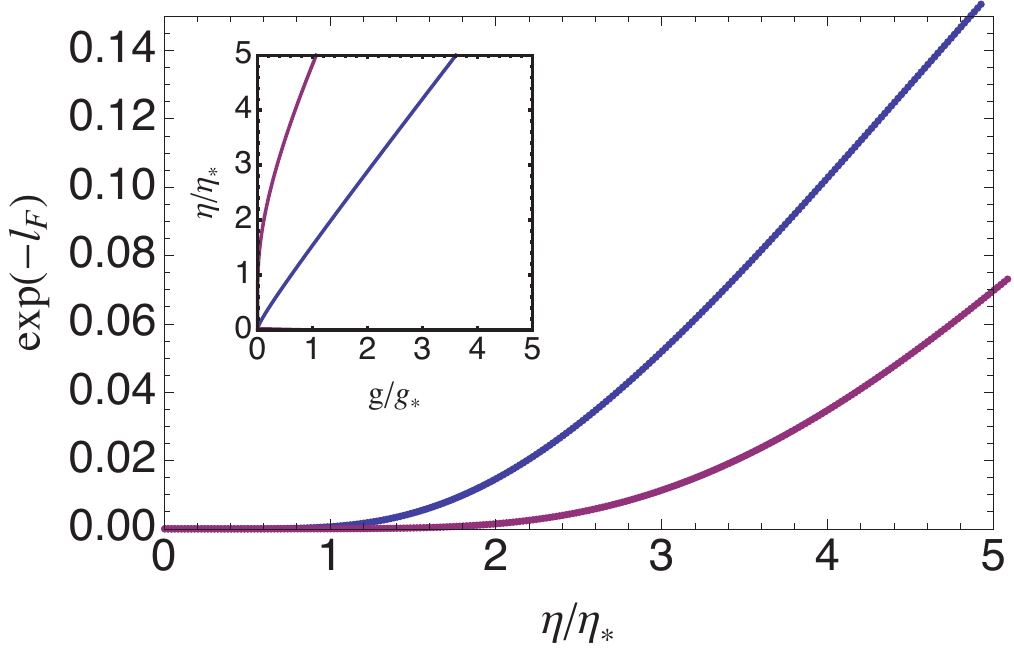}
\caption{The NFL energy scale $\exp(-l_F)$ calculated using Eq.~(\ref{eq:34}) along two lines in the $(\gz,\ez)$ plane (inset). Bottom line in inset is the pairing transition line, along which the NFL scale is top (blue) curve. Top line in inset gives the bottom (red) line for NFL scale.
The ratio $\gs/\es=1.22$.}
\label{fig:appenfl}
\end{figure}
The Fermi velocity flows to zero for any non-zero bare fermion-boson couplings (see Eq.~(\ref{eq:7})), but it does so in different ways depending on the bare couplings $(\gz,\ez)$.
We seek to identify two opposite regimes: (1) The ``NFL dominated'' regime where the typical NFL energy scale is much larger than the pairing gap energy; and (2) The ``pairing dominated'' regime where the converse is true. The former case  indicates\cite{Metlitski:2014p8103} that NFL behavior is observable at temperatures above the pairing (FQH) critical temperature. The typical NFL energy scale can be estimated as $E\sim\exp(-l_F)$ with $l_F$ the RG scale at which the Fermi velocity decays. In general the flow of $v_F$ (Eq.~(\ref{eq:7})) is given by $v_F/v_{F0}=\exp(-I_F(l))$, where we define
\begin{align}
  \label{eq:32}
  &I_F(l)\equiv\int_0^l(g(x)+\eta(x)) dx,\\\notag
  &I_F(l_F)\equiv1,
\end{align}
so that non-Fermi liquid effects become appreciable depending on the RG scale $l_F$.

In special case $\gs=\es\equiv\ees$, on the NQCL, the exact solution is
\begin{equation}
  \label{eq:33}
  l_F=\frac{1}{\ees}\ln\left(1+\frac{\ees(e-1)}{\gz+\ez}\right).
\end{equation}
On the line of fixed points, the exact solution for the scale of pairing (only happens for $\ez>\gz$) is $l_P=\frac{\pi}{2\sqrt{\ez-\gz}}$. Given that $\ees\ll1$, the NFL dominated regime appears close to the pairing transition at $\ez=\gz$, i.e., for $\ez-\gz\ll\ees^2$, since $l_P\gg l_F$ and both are $\gg1$. Conversely, the pairing dominated regime holds for $\ez-\gz\gg\ees^2$, where nematic coupling is much stronger than gauge coupling.

This analytic argument can be extended to stronger couplings, $\gz+\ez\gg\ees$, where $l_P\approx \frac{\pi}{2\sqrt{\ez-\gz}}$, and NFL dominated regime is found for $\ez-\gz\gg(\gz+\ez)^2$, while pairing dominated regime is found for $\ez-\gz\ll(\gz+\ez)^2$.

In the more general case $\ds=\es-\gs\neq0$ on the NQCL we get qualitatively the same results as above, which we use to sketch the dotted area in $(\gz,\ez)$ plane of Fig.~\ref{fig:1} denoting the regime $E\gg\Delta_P$.
For completeness, the exact expression for NFL scale here is:
\begin{equation}
  \label{eq:34}
  \frac{\gz}{\gs}e^{\gs l_F}+\frac{\ez}{\es}e^{\es l_F}=\frac{\gz}{\gs}+\frac{\ez}{\es}+e-1.
\end{equation}
In the limit $|\ds|\ll1$ the approximate solution is $E\sim\left(1+\frac{\es}{\gz+\ez}\right)^{-1/\es}$, qualitatively the same as Eq.~\ref{eq:33}.

It is clear from the $v_F$ Beta equation (Eq.~(\ref{eq:7})) that the NFL effects are enhanced by the ``total" fermion-boson coupling $\gz+\ez$. In the unfavorable limit of $\gz,\ez\rightarrow0$ the Fermi velocity takes on a logarithmically slow flow which makes the RG scale $l_F$ diverge and the NFL energy scale flat and nearly zero, see Fig.~\ref{fig:appenfl}.

\section{Flow in vicinity of NQCL}
\label{sec:appR}
\begin{figure}
  \centering
\includegraphics[width=0.49\textwidth]{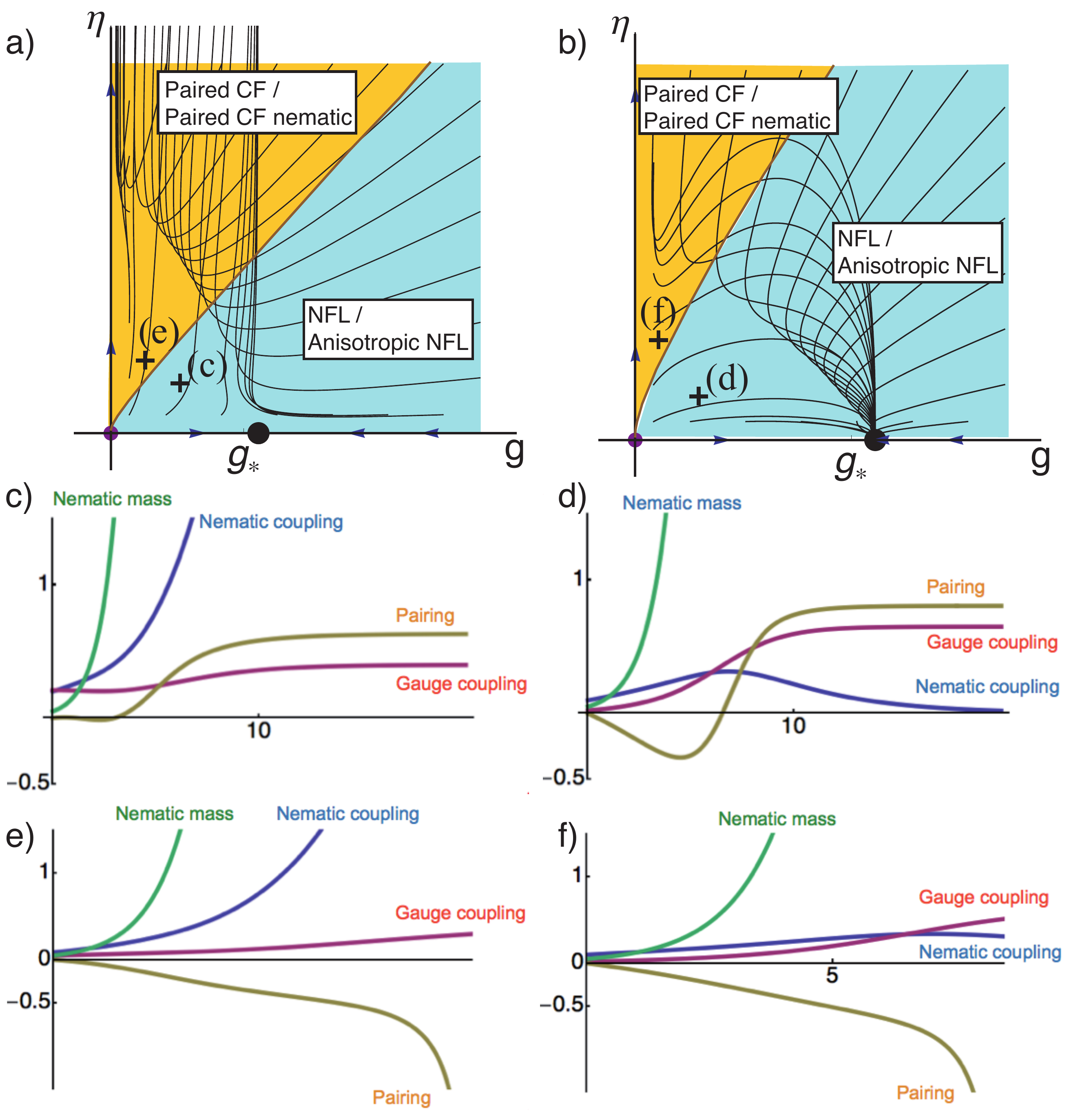}
\caption{Numerical RG flows projected onto fermion-boson coupling plane $(g,\eta$, in vicinity of the NQCL. Colored regions show $(\gz,\ez)$ values which give states corresponding to top[bottom] label when $r_0>0$[$r_0<0$]. In orange area there is infinitesimal pairing instability.
(a) Case $\epsilon_g<\epsilon_\eta$ corresponds to Fig.~\ref{fig:1}a, with $R_0/\gs=0.06$, $\gs/\es=8/13$.
(b) Case $\gs>\es$ corresponds to Fig.~\ref{fig:1}b, with $R_0/\gs=0.008$, $\gs/\es=13/8$. Here the topology of RG flows is the same as at the NQCL, except that the unstable fixed point $(0,\es)$ is replaced by $(0,\infty)$ affecting flow at $g=0$.
(c,d,e,f) Flows of $g$, $\eta$, $R$, and pairing $v$ in (c,e) correspond to initial values marked in panel (a); same for panels (d,f) marked in (b).
The horizontal axis is $\ln{l}$.
}
\label{fig:appptr}
\end{figure}
We define the dimensionless coupling constant $R\equiv r\kappa_\eta^2/\Lambda_y^{1+\epsilon_\eta}$ where by $r>0$ we label the mass of nematic fluctuations (see Eq.~\ref{eq:1b}): if $r_0>0$ then simply the $r=r_0$, while for $r_0<0$ in the vicinity of the nematic transition one has an action for the nematic fluctuations of the same form as Eq.~\ref{eq:1b}, with the bare mass of fluctuations positive at the new minimum and also labeled by variable $r>0$. Since $R$ is the ratio of the two terms quadratic in the nematic fluctuation field, it is relevant at any fixed point. The coupling of nematic and fermions remains strong near enough to the NQCL, where dimensionless mass $r v_F/u_\eta^2\Lambda_y<1$. In the case $R_0\neq0$, exact flows for $g,\eta$ can be given in integral form (see Fig.~\ref{fig:appptr}):
\begin{align}
  \label{eq:38}
  &\eta(l)=\frac{\ez \exp(\es l)}{1+p(l)-p(0)+F(l)},\\\notag
&g(l)=\frac{\gz \exp(\gs l)}{1+p(l)-p(0)+F(l)},\\\notag
  &p(l)=\frac{\gz}{\gs}e^{\gs l},\\\notag
  &F(l)=\ez\int_0^ldx\frac{1}{\exp(-\es x)+R_0 \exp(x/2)}.
\end{align}
For the case $\epsilon_g<\epsilon_\eta$ the most important feature of these flows compared to ones at the NQCL is that $F(l)$ replaces a term $\sim\exp(\es l)$. Since $F(l\rightarrow\infty)$ remains finite for any finite $R_0$, this implies that the $\eta$ diverges starting from any non-zero bare $\ez$. Analytical expressions for the $F$ can be obtained in limiting cases $\es=0$ and $\es=1$:
\begin{align}
  \label{eq:39}
&F(l,R_0,\es=0)=l-2\ln\left(\frac{1+R_0\exp(l/2)}{1+R_0}\right)\\\notag
&F(l,R_0,\es=1)=\frac{2}{\sqrt{R_0}}\left[\arctan(\sqrt{R_0}e^{l/2})-\arctan(\sqrt{R_0})\right]
\end{align}
with behavior
\begin{table}[h]
  \centering
  \begin{tabular}{lc|c}
    &$F(\es=0,l\rightarrow\infty)$&    $F(\es=1,l\rightarrow\infty)$\\
    \hline
    &$2\ln(1+1/R_0)$&$2/\sqrt{R_0}\left(\pi/2-\arctan(\sqrt{R_0})\right)$\\
    \hline
    $R_0\ll1$&$2|\ln(R_0)|$&$\pi/R_0$
  \end{tabular}
\end{table}

The fixed point $(0,\es)$ which is stable at NQCL in the case $\epsilon_g<\epsilon_\eta$ is therefore removed away from NQCL, and replaced by $(\gs,\infty)$, see Fig.~\ref{fig:appptr}a. Even though $\eta$ diverges the nematic fluctuations are suppressed by even faster growth of mass $R$, so the new fixed point is equivalent to a stable NFL fixed point at $(\gs,0)$. Consequently a transition line between the paired state and the NFL appears.
We now show how the transition line relates to the one for the case $\epsilon_g>\epsilon_\eta$ shown in Fig.~\ref{fig:appptr}b, and how both lines weakly depend on value $R_0\ll1$ compared to the transition line found at the NQCL in case $\epsilon_g>\epsilon_\eta$.
Following the simplified argument below Eq.~\ref{eq:43} (Appendix~\ref{sec:appg}), the pairing can occur if $f$ starts out positive (implying $\ez>\gz$) and the pairing RG scale $l'_T$ is estimated by the scale at which $f$ changes sign. Setting $f(l'_T)\equiv0$, we get:
\begin{equation}
  \label{eq:40}
  \frac{\ez}{\gz}=\exp(-\ds l'_T)+R_0\exp\left(\left(\frac{1}{2}+\gs\right)l'_T\right).
\end{equation}
In the relevant regime $\ez>\gz$, with $|\ds|\ll\es$ we obtain the following limits when $\ds<0$ (i.e. $\epsilon_g>\epsilon_\eta$)
\begin{equation}
  \label{eq:41}
l'_T=
\begin{cases}
    \frac{1}{1+\gs}\ln\left(\frac{\ez}{R_0\gz}\right),& R_0\gg b\\
        \frac{1}{|\ds|}\ln\left(\frac{\ez}{\gz}\right),& R_0\ll b
\end{cases},
\end{equation}
where we defined $b\equiv\left(\frac{\gz}{\ez}\right)^{(1/2+\es)/|\ds|}$. The latter limit connects to the NQCL and gives the same value for transition scale (Eq.~\ref{eq:12}), showing a very weak dependence on $R_0$ in this limit.
When $\epsilon_g<\epsilon_\eta$ (i.e. $\ds>0$) the limit $R_0\ll b$ is the one giving
\begin{align}
  \label{eq:42}
  &l'_T=\frac{1}{1+\gs}\ln\left(\frac{\ez}{R_0\gz}\right),
  &R_0\ll b,
\end{align}
so as expected for $\epsilon_g<\epsilon_\eta$ we find that the transition line exists only away from the NQCL ($R_0\neq0$).
These analytical results match well with numerical ones (Fig.~\ref{fig:app2opt}b), if one takes into account the caveats of numerical observation of pairing transition discussed below Eq.~\ref{eq:43} (Appendix~\ref{sec:appg}).

Another way of understanding the presence of a similar transition line behaviors for both $\epsilon_g\lessgtr\epsilon_\eta$ in vicinity of the NQCL is to focus on the general constraints in the pairing function $f(l)$. Away from the NQCL, it has the general form in Eq.~(\ref{eq:11}), and in stark contrast to case $R\equiv0$, where $f(l\rightarrow\infty)=\es-\gs$ it always has the limit $f(l\rightarrow\infty)=-\gs<0$ (no matter if $\eta$ diverges or not). So if $f(0)<0$, and $f(l)$ does not change sign, there is no possibility of a pairing instability. In general, analyzing Eq.~(\ref{eq:40}) as a function of $l'_T$ we can find the number of sign changes, which together with $f(0)=\ez/(1+R_0)-\gz$ and $f(\infty)=-\gs<0$ gives a qualitative picture of the pairing function $f(l)$. We find that $f(l)$ changes once from initially positive (promoting attraction) to negative, if $\ez/\gz>1+R_0$. In contrast it stays always negative if $\ez/\gz<m$, where $0<m<1+R_0$ is the minimum value of right hand side of Eq.~(\ref{eq:40}) as function of $l'_T$. Only when $\ds>R_0(1/2+\gs)$ does $m<1+R_0$, creating the possibility for $m<\ez/\gz<1+R_0$, for which $f(l)$ starts out and finishes negative, changing sign exactly twice.
The limit $R\rightarrow0$ is then obviously not universal, since apart from making $m=1$, it also allows values $f(\infty)$ that are not negative.

%

\end{document}